\documentclass[prl,superscriptaddress,twocolumn,showpacs,preprintnumbers,amsmath,amssymb]{revtex4}

\usepackage{graphicx}
\usepackage{bm}

\def\K{\,{\rm K}}
\def\cm{\,{\rm cm^{-1}}}
\def\ohmcm{\,\Omega^{-1}{\rm cm^{-1}}}
\def\mum{\,\mu{\rm m}}

\def\e{\epsilon}

\def\tc{T_{\rm c}}

\def\s1{\sigma_1(\omega)}
\def\e1{\epsilon_1(\omega)}

\begin{document}

\preprint{???}

\title{\boldmath  Superconducting energy gap and $c$-axis plasma frequency of (Nd,Sm)O$_{0.82}$F$_{0.18}$FeAs superconductors from 
infrared ellipsometry \unboldmath}

\author{A. Dubroka}
\author{K. W. Kim}
\author{M. R\"{o}ssle}
\author{V.K. Malik}%
 \affiliation{University of Fribourg, Department of Physics and Fribourg Center for Nanomaterials,
Chemin du Musee 3, CH-1700 Fribourg, Switzerland}

\author{R. H. Liu}

\author{G. Wu}

\author{X. H. Chen}
\affiliation{Hefei National Laboratory for Physical Sciences at Microscale and Department of Physics,
University of Science and Technology of China, Hefei, Anhui 230026, China}
\author{C. Bernhard}%
 \email{Christian.Bernhard@unifr.ch}

\affiliation{University of Fribourg, Department of Physics and Fribourg Center for Nanomaterials,
Chemin du Musee 3, CH-1700 Fribourg, Switzerland}

\date{\today}

\begin{abstract}
We present ellipsometric measurements of the far-infrared dielectric response of polycrystalline samples of the new pnictide superconductor {\it R}O$_{0.82}$F$_{0.18}$FeAs ({\it R}=Nd and Sm). We find evidence that the electronic properties are strongly anisotropic such that the optical spectra are dominated by the weakly conducting {\it c}-axis response similar as in the cuprate high-temperature superconductors (HTSC).
Accordingly, we obtain an upper limit of the {\it c}-axis superconducting plasma frequency of 
$\omega_{{\rm pl},c}^{\rm SC}\leq 260\cm$ which corresponds to a lower limit of the $c$-axis magnetic penetration depth of $\lambda_c\geq6\mum$ and an anisotropy of $\lambda_c/\lambda_{ab}\geq 30$ as compared to $\lambda_{ab}=185$\,nm from muon spin rotation [A. Drew {\it et al.}, cond-mat/0805.1042].  Also in analogy to the cuprate HTSC, our spectra exhibit the signatures of a gap-like suppression of the conductivity in the superconducting state with a large gap magnitude of 
$2\Delta\approx300\cm$ (37\,meV) and a ratio of $2\Delta/k_{\rm B}\tc\approx8$ that is suggestive of strong coupling.

\end{abstract}

\maketitle

The recent observation of superconductivity (SC) with critical temperatures, $\tc$, up to 55 K in the layered tetragonal pnictide {\it R}O$_{1-x}$F$_x$FeAs with {\it R} = La, Nd, Pr, Gd, and Sm marks the first discovery of a non copper-oxide-based layered high $\tc$ superconductor (HTSC) \cite{Kamihara1, Chen1, Ren1}. It raises the question whether a common pairing mechanism is responsible for HTSC in both the cuprates and the pnictides. 
Similar like the cuprates, the pnictides have a layered structure that is comprised of alternating FeAs and LaO sheets with Fe arranged on a square lattice~\cite{Kamihara1}. Theoretical calculations predict a quasi two-dimensional electronic structure with metallic FeAs layers and LaO layers that mainly act as blocking layers and as charge reservoir upon chemical substitution \cite{Lebeque07, Singh1, Haule1}. Also in analogy to the cuprates, SC emerges upon doping away from a magnetic mother compound, the maximal $\tc$ occurring just as magnetism disappears \cite{Chen2,Mook1,Mandrus1,Klauss1}.
There are also some clear differences with respect to the cuprates. Band structure calculations suggest that the pnictides are multiband superconductors with up to five FeAs-related bands crossing the Fermi-level \cite{Lebeque07,Singh1,Haule1,Xu1} as opposed to the cuprates which, due a strong Jahn-Teller distortion, have only one relevant 
Cu(3$d_{x^2-y^2}$)O band. Furthermore, in these pnictides the highest $\tc$ values are achieved upon electron doping and not for hole doping \cite{Wen1} as in the cuprates \cite{Luke90}.

Further progress in assessing the differences and similarities of these cuprate and pnictide superconductors requires experimental information especially about their electromagnetic properties. The research into the cuprate HTSC has shown that infrared spectroscopy can play an important role since it provides fairly direct and reliable information about the electronic properties in the normal state as well as in the SC state~\cite{Tanner92, Basov05}. Even measurements on polycrystalline samples yielded first important information. In particular, it was established that thanks to the very large electronic anisotropy of the cuprates, the pronounced features of the reflectivity spectra are representative of the weakly conducting $c$-axis response. The metallic $ab$-plane component merely gives rise to a moderate increase of the overall magnitude of the conductivity with respect to the one of $c$-axis component~\cite{Schlesinger87,Tanner92,Uchida96,Basov05,Bonn87,Bonn87YBCO}. This interpretation has been confirmed by direct measurements of the in-plane and $c$-axis response on single crystals \cite{Uchida96,Basov05}. Accordingly, measurements on polycrystals can provide reliable information, for example on the eigenfrequency of the $c$-axis phonon modes, the upper limit of the $c$-axis plasma frequency of the SC condensate, $\omega_{\rm pl}^{\rm SC}$, and the magnitude of the SC energy gap, $\Delta$~\cite{Schlesinger87,Tanner92,Uchida96,Basov05,Bonn87,Bonn87YBCO}. 
 
Concerning the optical properties of the pnictides, so far only few reports have been reported on the undoped mother compound \cite{Wang1} and superconducting LaO$_{0.9}$F$_{0.1}$FeAs \cite{Wang2,Drechsler1} which did not detail the impact of SC on the dielectric function. 

In this letter we present ellipsometric measurements of the far-infrared dielectric response of polycrystalline samples of {\it R}O$_{0.82}$F$_{0.18}$FeAs with {\it R} = Nd and Sm and 
$\tc= 52(2)$ and 45(3) K, respectively. In the first place, our data reveal that the electronic properties are strongly anisotropic, similar to the cuprate HTSC. From our data we thus obtain an upper limit of the $c$-axis SC plasma frequency of $\omega_{{\rm pl},c}^{\rm SC}\leq260\cm$ and a magnetic penetration depth of 
 $\lambda_c\geq6\mum$, compared with the in-plane values of $\lambda_{ab}=185$\,nm  and $\omega_{{\rm pl},ab}^{\rm SC}\approx8380\cm$ from muon spin rotation measurements \cite{Luetkens1,Drew1}.
In addition, our data show that the magnitude of the SC energy gap is $2\Delta=300\cm$ with a ratio of $2\Delta/k_{\rm B}\tc\approx8$ typical of strong coupling SC.

\begin{figure}
\includegraphics[width=6cm]{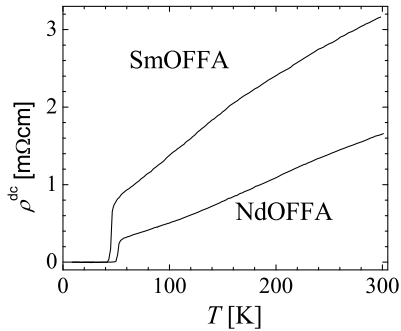}
\caption{\label{Trans} Temperature dependent resistivity of the polycrystalline 
NdO$_{0.82}$F$_{0.18}$FeAs and SmO$_{0.82}$F$_{0.18}$FeAs samples.}
\end{figure}

Polycrystalline samples with nominal composition NdO$_{0.82}$F$_{0.18}$FeAs (NdOFFA) and SmO$_{0.82}$F$_{0.18}$FeAs  (SmOFFA) have been synthesized by conventional solid state reaction methods as described in Refs.~\cite{Chen1,Chen2}. Standard powder x-ray diffraction patterns were measured where all peaks could be indexed to the tetragonal ZrCuSiAs-type structure. dc resistivity (see Fig.~\ref{Trans}) and magnetisation measurements were made to determine the midpoint (10\% to 90\% width) of the resistive and diamagnetic transitions $\tc$ ($\Delta\tc$) of 52(3) K for NdOFFA of 45(3) for SmOFFA. The samples were polished using diamond suspension to obtain flat and shiny surfaces. While the surfaces were certainly not perfectly mirror-like, the material had a high density and did not give rise to significant depolarisation effects as confirmed by UV ellipsometry measurements.  

The infrared ellipsometry measurements in the range 45 to 640 $\cm$ (5 - 80 meV) were performed with a home-built setup attached to a Bruker 113V Fast-Fourier spectrometer as described in Ref.~\cite{Bernhard04}. The angle of incidence of the polarised light was 80$^\circ$. 
Ellipsometry enables one to directly measure the complex dielectric function, 
$\tilde{\epsilon}(\omega)=\epsilon_1(\omega)+{\rm i} \epsilon_2(\omega)$, and the related optical conductivity 
$\tilde{\sigma}(\omega)=-{\rm i}\omega\epsilon_0(\tilde{\epsilon}(\omega)-1)$, without a need for Kramers-Kronig analysis \cite{Azzam77}. Furthermore, it is a self-normalizing technique that enables very accurate and reproducible measurements. 

\begin{figure*}
\includegraphics[width=16cm]{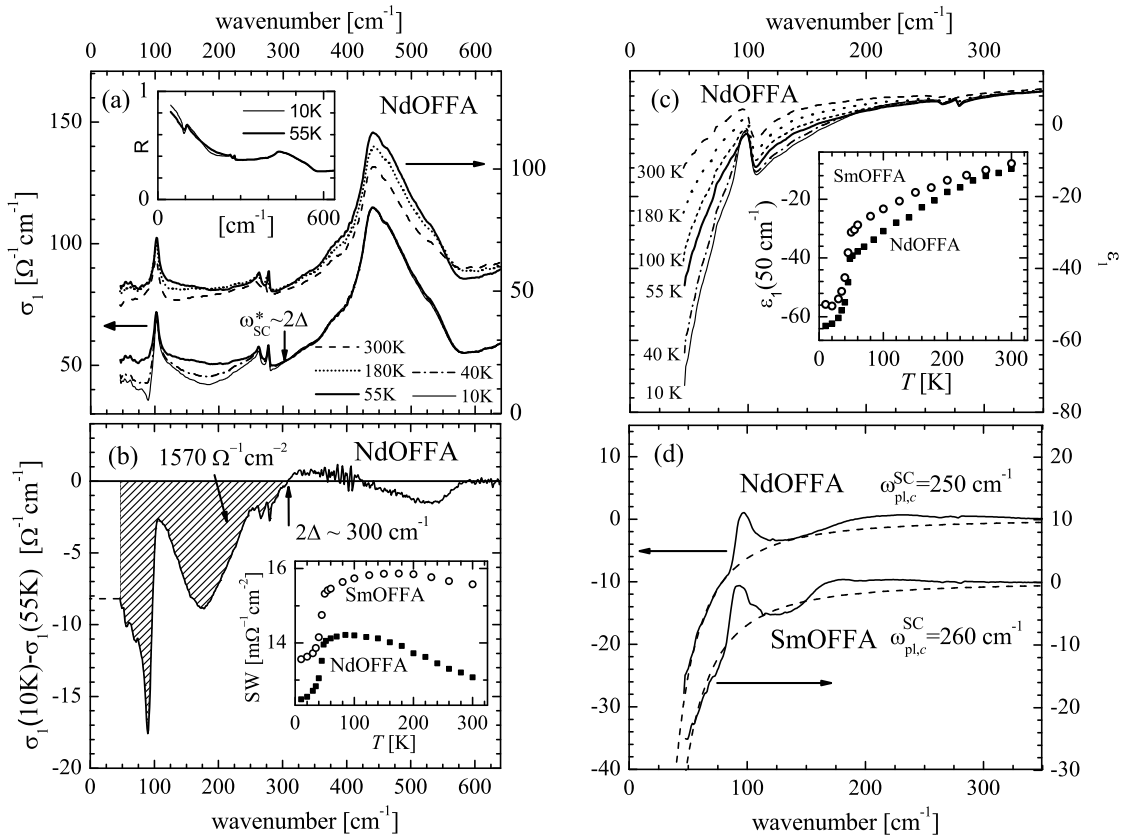}
\caption{\label{Infra} Temperature dependence of the infrared dielectric response of the
(Nd,Sm)O$_{0.82}$F$_{0.18}$FeAs superconductors. (a)~Representative spectra of the real part of the conductivity, $\s1$. The arrow marks the onset of the SC induced gap-like suppression. The inset shows calculated reflectivity spectra. (b)~Difference between the conductivity in the normal and the SC state in terms of $\sigma_1(10\K)-\sigma_1(55\K)$. The shaded area indicates the missing spectral weight due to the SC condensate. The inset details the temperature dependence of the integral of $\s1$ between 45 and $300\cm$ (SW). (c)~Corresponding spectra of the real part of the dielectric function, $\e1$. The inset details the temperature evolution of $\epsilon_1(50\cm)$. (d)~Difference spectra  $\epsilon_1(10\K)-\epsilon_1(\tc)$ [solid lines]
showing the SC induced change of $\e1$. Dotted lines show fits with the function 
$-(\omega^{\rm SC}_{{\rm pl},c}/\omega)^2$.}
\end{figure*}

Figure~\ref{Infra} displays the infrared optical spectra of the NdOFFA sample at representative temperatures between 10 and 300 K. Figure~\ref{Infra}(a) shows the real part of the optical conductivity, $\s1$, and 
Fig.~\ref{Infra}(c) the corresponding real part of the dielectric function, $\e1$, while Figs.~\ref{Infra}(b) and~\ref{Infra}(d) detail the SC induced changes. The inset of Fig.~\ref{Infra}(a) shows the calculated reflectivity spectra which agree well with previously reported ones~\cite{Wang1,Drechsler1}. 
It is immediately evident from our data that the electronic part of the optical response is extremely weak. The most pronounced features are indeed due to the infrared-active phonon modes which give rise to pronounced, narrow peaks near 102, 257, 270 and 440 $\cm$. The electronic part of $\s1$ has a surprisingly small magnitude of less than 100 $\ohmcm$ and there is only a very weak signature of an inductive response in $\e1$, which becomes negative only below 300$\cm$. We note that very similar optical data have been obtained on four corresponding samples which contained Sm and La instead of Nd.

This weakly conducting behaviour needs to be reconciled with the metallic dc transport with $\rho^{\rm dc}=0.35\,{\rm m\Omega cm}$ and $\sigma^{\rm dc}=2800\ohmcm$ 
at 60~K (the NdOFFA sample) and also with the short in-plane magnetic penetration depth of $\lambda_{ab}=185-235$~nm and thus sizeable SC plasma frequency of 
$\omega^{\rm SC}_{{\rm pl},ab}\approx 6500-8500 \cm$ as obtained 
from muon spin rotation (in parts on the same samples) \cite{Luetkens1,Drew1}. 
The explanation of this puzzling behaviour can be found in the literature on the earlier infrared studies on polycrystalline samples of the cuprate 
HTSC~\cite{Schlesinger87, Bonn87, Takagi89}. Very similar trends were observed here, that are meanwhile well understood in terms of the very strong electronic anisotropy of the electronic transport parallel and perpendicular to the conducting CuO$_2$ planes which are separated by various kinds of essentially insulating blocking layers. Notably, it was found that the infrared spectra on polycrystalline samples are dominated by the characteristic features of the nearly insulating $c$-axis response. The overall conductivity of the electronic background was found~\cite{Schlesinger87} to be much lower than the dc conductivity as measured by transport. For example in La$_{2-x}$Sr$_x$CuO$_4$~\cite{Bonn87} the dc extrapolation of $\s1$ right above $\tc$ yields values of at most $200\ohmcm$, whereas transport measurements on polycrystals give dc conductivities of $10^3-10^4\ohmcm$ \cite{Takagi89}. To the contrary, for isotropic crystalline materials like the manganite perovskites, it is well known that the dc conductivity inferred from infrared data on polycrystals is systematically higher than the corresponding values from transport measurements which are decreased by grain boundary effects~\cite{Kim97}. 
The previous works on polycrystalline \cite{Bonn87, Bonn87YBCO}  and single crystalline samples ~\cite{Uchida96, Basov05} of the cuprate HTSC have also shown that many aspects of 
the far-infrared $c$-axis dielectric function can be reliably deduced from measurements on polycrystalline samples. This includes besides the infrared active phonon modes, the magnitude of the SC energy gap, and the upper limit of the plasma frequency of the $c$-axis SC condensate~\cite{Tanner92}.  

Returning to the pnictide superconductors, our infrared data thus highlight a very strong anisotropy of the electronic responses parallel and perpendicular to the FeAs layers. Furthermore, they enable us to extract important information about the $c$-axis response of these new superconductors. As outlined below, we can obtain the magnitude of the SC energy gap and also a reliable lower limit of the anisotropy of the SC magnetic penetration depth of $\lambda_c/\lambda_{ab}\geq 30$. 

The magnitude of the SC energy gap is indeed apparent in 
Fig.~\ref{Infra}(a) due to the well-resolved gap-like suppression of $\s1$ below the onset frequency of $\omega^*_{\rm SC}\approx300\cm$ which is detailed in Fig.~\ref{Infra}(b). Previous similar studies on the cuprate HTSC have shown that this onset frequency (in the relevant so-called dirty limit of the weakly conducting {\it c}-axis response) approximately scales as twice the maximum value of the SC gap energy, i.e., $\omega^*_{\rm SC}\approx2\Delta$~\cite{Tinkhambook,Tanner92,Hirschfeld97,Palumbo96}. Accordingly, we estimate $\Delta\approx19$\,meV and $2\Delta/k_{\rm B}\tc\approx8$ 
which is considerably larger than the value of 3.5 predicted by standard weak-coupling BCS theory. We note that such a large gap magnitude is consistent with the very high upper critical fields as derived from recent magneto-transport measurements~\cite{Zhu1}. Furthermore, we remark that similarly large gap magnitudes and upper critical fields occur in the cuprate HTSC where it has been argued that they provide evidence for a strong-coupling SC pairing mechanism. 

The shape of the gap is somewhat obscured by the anomalous temperature dependence of the phonon mode near $100\cm$ which develops an asymmetric shape below $\tc$ [see Figs.~\ref{Infra}(a) and ~\ref{Infra}(b)]. 
This shape change gives rise to the wavy structure in the conductivity difference spectrum in Fig.~\ref{Infra}(b). Such behaviour might be taken as evidence that this phonon mode strongly couples to the electronic background. However, it appears that the peak frequency of the phonon mode does not exhibit any anomalous behaviour. Furthermore, we remark that similar anomalous shapes of phonon modes were previously observed on polycrystalline cuprate HTSC samples~\cite{Bonn87YBCO} that were not confirmed by subsequent measurements on single crystals~\cite{Basov05}. 

As already mentioned above, our optical data also allow us to access another important parameter of the SC state which are the $c$-axis components of the  plasma frequency of the SC condensate, $\omega^{\rm SC}_{{\rm pl},c}$ and the related magnetic penetration depth, $\lambda_c$. These values can be derived from our data in two independent ways (since ellipsometry measures $\s1$ and $\e1$ independently). Firstly, one can obtain them from the so-called missing area due to the gap-like suppression in the regular part of $\s1$. As shown in Fig.~\ref{Infra}(b) by the shaded area for the measured frequency range of $45-300\cm$, this amounts to $1570\ \Omega^{-1}{\rm cm}^{-2}$ or $\omega^{\rm SC}_{{\rm pl},c}=245\cm$. These values are likely even  somewhat higher due to a contribution from the frequency range below $45\cm$. As an example, the dotted line shows an straight extrapolation which would yield a value of $1940\ \Omega^{-1}{\rm cm}^{-2}$ or $\omega^{\rm SC}_{{\rm pl},c}=270\cm$. 
Secondly, the plasma frequency of the SC condensate
can be determined from the inductive response in $\e1$ where a SC induced contribution is apparent in the temperature evolution of $\epsilon(50\cm)$, see the inset of Fig.~\ref{Infra}(c). 
As shown in Fig.~\ref{Infra}(d), reasonable fits to the low frequency parts of the difference spectra of $\e1$ between the 10~K and the temperature right above $\tc$, can be obtained with the function, 
$-(\omega^{\rm SC}_{{\rm pl},c}/\omega)^2$, that accounts for the inductive response due to the SC condensate. 
For both samples we obtained a similar value of $\omega^{\rm SC}_{{\rm pl},c}=250-260\cm$ which translates into a spectral weight of the SC condensate of 
$\approx1700\ \Omega^{-1}{\rm cm}^{-2}$. 
We note, that the determination of $\omega^{\rm SC}_{{\rm pl},c}$ in both cases is based on the assumption that the regular part of the dielectric function does not change between 10 and 55~K. Due to a possible narrowing of the regular part below $\tc$, our analysis may well give a slightly overestimated value of $\omega^{\rm SC}_{{\rm pl},c}$ which nevertheless provides a reliable upper boundary. Accordingly, we can derive from our data a lower limit of the magnetic penetration depth in the 
$c$-axis direction of $\lambda_c\geq 6\mum$. With the in-plane values of $\lambda_{ab}\approx 185-235$ nm as obtained from recent muon-spin-rotation measurements (that were in parts performed on the same samples) \cite{Luetkens1, Drew1} this yields an anisotropy of $\lambda_c/\lambda_{ab}\geq 30$. 
It is interesting to note that rather similar values have been obtained for weakly underdoped to optimally doped La$_{2-x}$Sr$_x$CuO$_4$ single crystals~\cite{Shibauchi94}.

In summary, we reported infrared optical measurements of polycrystalline samples of the 
(Nd,Sm)O$_{0.82}$F$_{0.18}$FeAs superconductors. We outlined that the optical spectra provide evidence for a strong electronic anisotropy and a very weakly conducting $c$-axis response, similar like in the cuprate HTSC. We deduced important parameters like a lower limit for the $c$-axis magnetic penetration of 
$\lambda_c\geq6\mum$
and an anisotropy of $\lambda_c/\lambda_{ab}\geq 30$  compared to $\lambda_{ab}=185$~nm from muon spin rotation~\cite{Drew1}. We also determined the gap magnitude of 
$2\Delta\approx300\cm$ and the ratio of $2\Delta/k_{\rm B}\tc\approx8$ that significantly exceeds the value of 3.5 as predicted by weak-coupling BCS theory. Overall our measurements reveal several similarities between the optical spectra of the new pnictide superconductors and the ones of the cuprate HTSC. 

Acknowledgment: This work is supported by the Schweizer Nationalfonds (SNF) with grant 200020-119784 and by the Deutsche Forschungsgemeinschaft (DFG) with grant BE2684/1-3 in FOR538.

\end{document}